# Methodology for Calculating CO2 Absorption by Tree Planting for Greening Projects

**Kento Ichii[1], Toshiki Muraoka[2], Nobumichi Shinohara[1], Shunsuke Managi[1], and Shutaro Takeda[1]**

[1] Urban Institute, Kyushu University
[2] Graduate School of Advanced Integrated Studies in Human Survivability, Kyoto University

## Abstract

In order to explore the possibility of carbon credits for greening projects, which play an important role in climate change mitigation, this paper examines a formula for estimating the amount of carbon fixation for greening activities in urban areas through tree planting. The usefulness of the formula studied was examined by conducting calculations based on actual data through measurements made by on-site surveys of a greening companie. A series of calculation results suggest that this formula may be useful. Recognizing carbon credits for green businesses for the carbon sequestration of their projects is an important incentive not only as part of environmental improvement and climate change action, but also to improve the health and well-being of local communities and to generate economic benefits. This study is a pioneering exploration of the methodology.

## 1. Introduction

### *1.1 Carbon credit*

Carbon credits are being viewed by governments and businesses around the world as an important tool for reducing greenhouse gas emissions. Carbon credits provide incentives for companies and countries to reduce greenhouse gas emissions. Not only can they slow the progression of climate change by investing in emission reduction projects, carbon credits are a market-baszwed mechanism that provides economic benefits to companies and organizations that achieve emission reductions. This facilitates the transition to sustainable business models, and carbon credit markets standardize the measurement, reporting, and verification of emission reductions, providing transparency and credibility. For these reasons, carbon credits play a vital role in the fight against global climate change.

These mechanisms include carbon taxes, cap-and-trade systems, and hybrid approaches. The effectiveness of carbon pricing mechanisms in reducing emissions and stimulating innovation has been studied (Emeka-Okoli et al., 2024). However, challenges remain, including uneven geographic dispersion of investments and untapped abatement opportunities in some sectors (Larson et al. ., 2008). Future research should focus on optimizing carbon trading operations, addressing limitations, and exploring innovative approaches to enhance the effectiveness of these mechanisms.

Among the mechanisms, the J-Credit program certifies the amount of greenhouse gas emission reductions and removals in Japan and makes them tradable as credits. The J-Credit Program selects projects that contribute to the reduction or absorption of greenhouse gas emissions (Fukushima, 2022). This includes projects that introduce renewable energy, promote energy conservation, and absorb forests. For forest absorption projects, the amount of carbon dioxide absorbed through the planting of new forests or the conservation of existing forests will be measured, and credits will be awarded based on this amount. For energy conservation projects, credits are awarded based on the amount of energy consumption reduced through the introduction of energy conservation equipment and operational improvements. For renewable energy projects such as wind power and solar power, the credit is the difference between the amount of electricity generated by the project and the amount of emissions that would have been generated if the same amount of electricity had been supplied by conventional fossil fuel power generation. Through these processes and methods, the J-Credit system enables the issuance and trading of highly reliable carbon credits, thereby contributing to the reduction of greenhouse gas emissions.

### *1.2 Greening projects and carbon*

On the other hand, despite playing a major role, greening projects are not currently included in these carbon credit frameworks. Greening projects refer to activities aimed at increasing the number of plants by planting trees and developing green spaces in various areas, including urban, rural, and forested areas. These projects have a wide range of objectives, including environmental protection, climate change mitigation, biodiversity conservation, urban beautification, and improvement of residents' health. For urban street tree planting, planting trees along roads and in parks beautifies the urban landscape, absorbs carbon dioxide, and reduces air pollution. In addition, the development of urban parks provides residents with parks as places to relax, promote health, and support the formation of communities. Greenways and green corridors will be developed to provide green corridors that link the city and nature, and provide comfortable spaces for pedestrians and cyclists to move around.

If these greening projects, which play an important role in mitigating climate change, are certified as carbon credits in the future, they will be an important incentive not only to improve the environment and combat climate change, but also to improve the health and welfare of local communities and generate economic benefits. This paper presents a pilot formula to calculate the amount of carbon fixation by this greening project and discusses its inclusion in future carbon credits.

## 2. Method

### *2.1 Basic assumptions*

In this paper, the calculation formula will be calculated for activities aimed at increasing the number of plants, especially in urban areas through the development of green spaces by planting trees. In particular, in the calculations, carbon fixation will be calculated for sales results related to tall trees, medium-sized trees, and shrubs. Based on the estimated height, diameter, and survival rate for all periods after tree planting, and the increase in trunk volume, the amount of CO2 absorbed by trees during a given period is calculated.

The project period will be 100 years, and the amount of CO2 emissions generated in the implementation of the project will be subtracted from the amount of CO2 absorption contribution, which will be the final amount of credits generated. The distribution of the contribution among stakeholders, such as nursery tree distributors, greening business operators, general contractors, etc., during the 100-year period will be determined on a case-by-case basis.

### *2.2 Proposed formula*

The following formula (1) can be used to calculate the amount of CO2 absorbed per tree over 100 years, taking into account the possibility that trees will continue to be planted without being removed.

CO2 absorption per tree over 100 years

$$= \int_0^{99} (1-p)^t pH(t)\left(\frac{d(t)}{2}\right)^2 \pi c dt + (1-p)^{100} H(100)\left(\frac{d(100)}{2}\right)^2 \pi c$$

... Eq. (1)

Of which, amount of CO2 absorbed that can be recognized as credits

$$= (1-p)^{100} H(100)\left(\frac{d(100)}{2}\right)^2 \pi c$$

... Eq. (2)

WHEREAS,

Assume p for the likelihood that the tree will be removed during the year.

Let H(t) [cm] be the height of the tree t years after it was first planted.

Let d(t) [cm] be the diameter of the tree trunk.

$\pi$ is pi.

The remaining parameters used in this analysis to calculate CO2 absorption are constants, and are therefore assumed to be c [t-C./ cm3 ].

The likelihood that the trees will still be planted t years after they were first planted, $(1-p)^t$ is

In this formula, the first term expresses the amount of CO2 absorbed from planting to that point and its probability of occurrence if the tree is done after t years. The second term expresses the amount of CO2 absorbed and the probability of its occurrence that the tree will continue to be planted after 100 years from planting.

### *2.3 Probability of tree removal*

Referring to the "Street Trees in Japan IX" published by the National Institute for Land and Infrastructure Management (NILIM) of the Ministry of Land, Infrastructure, Transport and Tourism, we calculate the likelihood p that trees will be removed due to felling or falling within a year.

As shown in the following figures on the number of tall trees nationwide and the number of medium-sized shrubs nationwide, the number of street trees, which had increased rapidly since 1987, remained flat from 2007 to 2017 for both tall trees and medium-sized shrubs. This transition indicates a shift to a steady state, with no major changes in new planting or removal policies occurring since 2007.

At the same time, this means that the number of newly planted street trees was equal to the number of street trees removed due to felling or falling trees. Assuming that the average life span of street trees is 35 years and that 1/35 of the total number of trees is removed each year, the number of new tall trees planted per year is calculated to be 6.67 million trees ÷ 35 years = 190.57 million trees/year.

Assuming the same conditions continued during the period 2007-2022, the total number of new trees planted would have been 190,570,000 trees/year × 15 years = 285.857 million trees. The total number of street trees removed due to downed trees is estimated to be 380,000 + 285.857 million = 323.857 million trees.

The total number of street trees that were in a state of being planted after 2007 is 9.528.57 million, which is the sum of 6.67 million trees that were planted as of 2007 and 2.855.857 million newly planted over the 15-year period. Therefore, during the 15-year period from 2007 to 2022, 3.238.570 million trees / 9.952.857 million trees = 33.988% of the street trees were removed. This means that over the 15-year period from 2007 to 2022, an average of about 2.7309% per year ($(1-(1-0.33988)^{1/15} \approx 0.027309)$ of street trees will be removed. Thus p is 0.027309 (2.7309%). Note that if on average about 2.7309% of street trees are removed per year, the average life span of a street tree is 36.12 ($\int_0^\infty (1-0.027309)^t dt = 36.12$) years.

Similarly, for medium and low trees, assuming that the average life span of street trees is 35 years and that 1/35 of the total number of trees is removed each year, the number of new tall trees planted per year is calculated to be 139.79 million trees ÷ 35 years = 3.994 million trees/year.

Assuming that the same conditions continued from 2007 to 2022, 3,994,000 new trees were planted per year × 15 years = 59.91 million new trees were planted. The total number of street trees removed due to tree felling or toppling over the 15-year period from 2007 to 2022 is estimated to be 59.91 million + 4.65 million = 64.56 million.

The total number of street trees that remained planted after 2007 is 199.7 million, which is the sum of the 139.79 million trees planted as of 2007 and 59.91 million newly planted over the 15-year period. Therefore, during the 15-year period from 2007 to 2022, 64.56 million trees / 199.7 million trees = 32.32849% of the street trees were removed. This translates to an average annual removal rate of approximately 2.56977% ($1-(1-0.3232849)^{1/15} \approx 0.0256977$) of street trees will be removed. Thus p is 0.0256977 (2.56977%). Note that if on average about 2.56977% of street trees are removed per year, the average life span of street trees is 38.41 ($\int_0^\infty (1-0.0256977)^t dt = 38.41$) years.

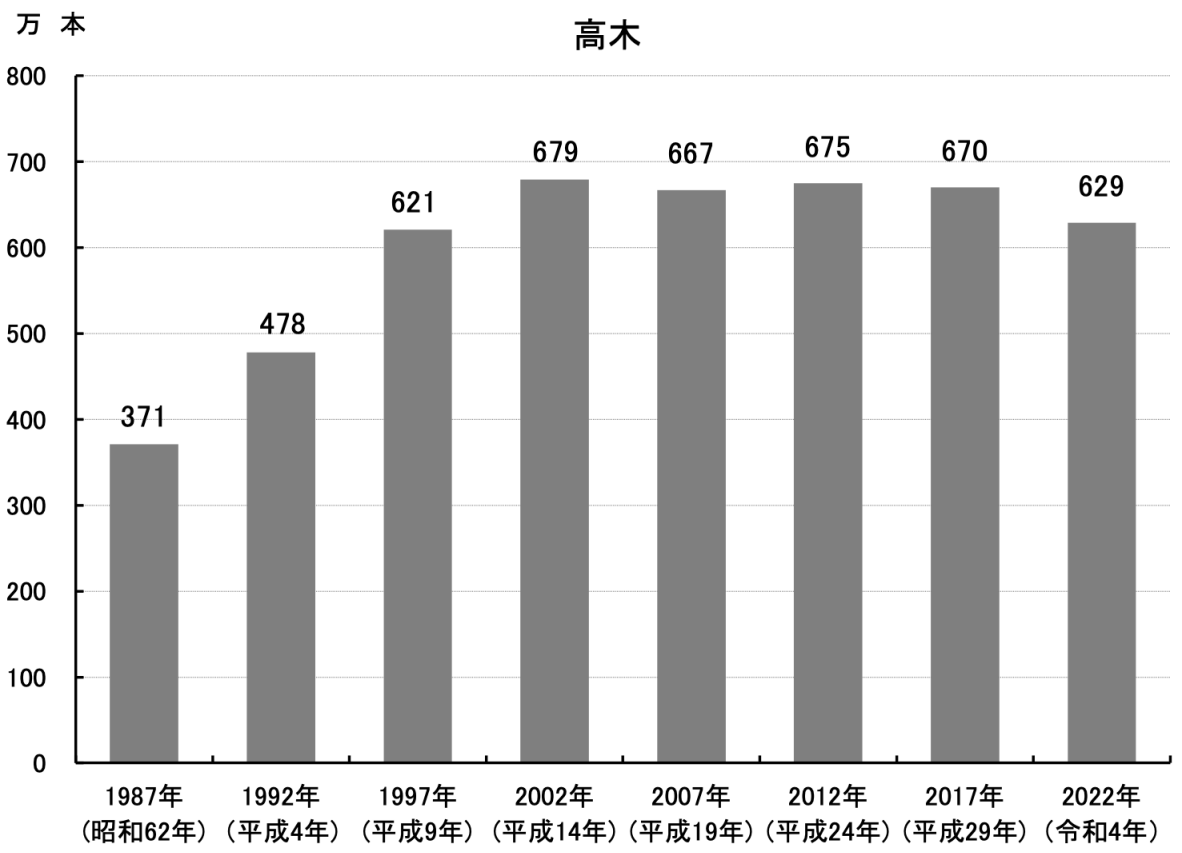

Figure 1 Number of tall trees in Japan

From "Street Trees in Japan IX" published by the National Institute for Land and Infrastructure Management (NILIM) of the Ministry of Land, Infrastructure, Transport and Tourism.

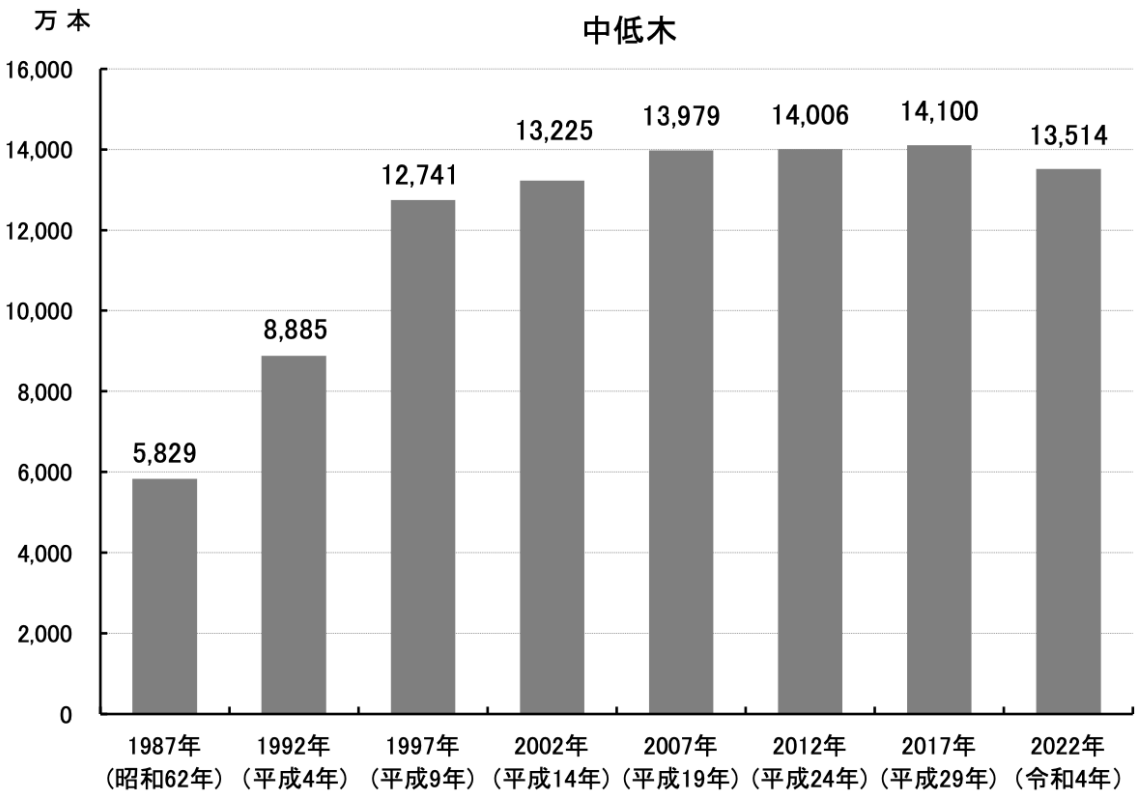

Figure 2 Number of medium-sized shrubs in Japan

From "Street Trees in Japan IX" published by the National Institute for Land and Infrastructure Management (NILIM) of the Ministry of Land, Infrastructure, Transport and Tourism.

### *2.4 Calculation of tree height H(t)*

Referring to the "Tree Growth Model" published by the Ecological Society of Japan, calculate the tree height H(t)(cm) after t years of planting of tall, medium-sized trees and shrubs of evergreen, deciduous and coniferous trees.

H(t) for each tree species is obtained from the following formula.

Evergreen (tall tree)

$$H(t) = 2500(1-0.975^t)$$

Deciduous tree (tall tree)

$$H(t) = 2500(1-0.962^t)$$

Conifers (tall trees)

H(t) = $35 + 5471(1 - e^{-0.00592(t-1)})^{0.65669}$

Evergreen (medium tree)(t<16.412)
H(t) = $2500(1 - 0.975^{t})$

Evergreen (medium tree) (t≥16.412)
H(t) = 850

Deciduous tree (medium tree) (t<16.412)
H(t) = $2500(1 - 0.962^{t})$

Deciduous tree (medium tree) (t≥16.412)
H(t) = 850

Conifer (medium tree)(t<16.412)
H(t) = $35 + 5471(1 - e^{-0.00592(t-1)})^{0.65669}$

Conifer (medium wood) (t≥16.412)
H(t) = 850

Shrub (t<3.72093)
H(t) = $107.5t$

Shrub (t≥3.72093)
H(t) = 400

## *2.5 Calculation of tree diameter d(t)*

Referring to the "Data Relating Height and Trunk Girth by Tree Type" published by the Ministry of Land, Infrastructure, Transport and Tourism, the diameter of evergreen trees, deciduous trees, and coniferous trees is calculated in relation to tree height H(t).

The relationship between tree height and diameter is shown below, based on the "Tree Height and Trunk Girth Relationship Data by Species" published by the Ministry of Land, Infrastructure, Transport and Tourism.

### *2.5.1 Evergreen tree*

The calculation table for evergreen tree is as follows:

Table 1 Evergreen tree diameter vs. height

| Tree height (cm)(=H) | Trunk circumference (cm) | Diameter (cm) (=d) |
|---|---|---|
| 250 | 11 | 3.503185 |
| 300 | 16 | 5.095541 |
| 350 | 20 | 6.369427 |
| 400 | 37 | 11.78344 |
| 450 | 36 | 11.46497 |
| 500 | 35 | 11.1465 |
| 550 | 52 | 16.56051 |
| 600 | 64 | 20.38217 |
| 650 | 70 | 22.29299 |
| 700 | 83 | 26.43312 |
| 800 | 95 | 30.25478 |
| 850 | 100 | 31.84713 |
| 1000 | 120 | 38.21656 |
| 1100 | 150 | 47.7707 |

### *2.5.2 Deciduous tree*

The calculation table for deciduous tree is as follows:

Table 2 Deciduous tree diameter vs. height

| Tree height (cm)(=H) | Trunk circumference (cm) | Diameter (cm) (=d) |
|---|---|---|
| 200 | 10 | 3.184713 |
| 250 | 11 | 3.503185 |
| 300 | 13 | 4.140127 |
| 350 | 18 | 5.732484 |
| 400 | 22 | 7.006369 |
| 450 | 26 | 8.280255 |
| 500 | 32 | 10.19108 |
| 550 | 35 | 11.1465 |
| 600 | 43 | 13.69427 |
| 650 | 55 | 17.51592 |
| 700 | 54 | 17.19745 |
| 800 | 83 | 26.43312 |
| 900 | 83 | 26.43312 |
| 1000 | 107 | 34.07643 |
| 1100 | 135 | 42.99363 |

### *2.5.3 Conifer*

The calculation table for conifer tree is as follows:

Table 3 Conifer tree diameter vs. height

| Tree height (cm)(=H) | Trunk circumference (cm) | Diameter (cm) (=d) |
|---|---|---|
| 250 | 13 | 4.140127 |
| 300 | 15 | 4.77707 |
| 350 | 19 | 6.050955 |
| 400 | 23 | 7.324841 |
| 450 | 26 | 8.280255 |
| 500 | 33 | 10.50955 |
| 550 | 40 | 12.73885 |
| 600 | 43 | 13.69427 |
| 700 | 53 | 16.87898 |
| 800 | 60 | 19.10828 |
| 900 | 80 | 25.47771 |
| 1100 | 100 | 31.84713 |

### *2.6 Calculation of constant c*

Among the parameters included in the calculation of CO2 absorption, prepared by referring to the "Visualization Calculation Sheet" published by the Forestry Economic Research Institute, "BEF x (1+RtSR) x BD x CF x (44/12) x $10^{-6}$ is a constant, so it is assumed to be c [t-C./ cm3].

WHEREAS,
BEF [no units] is the Biomass Expansion Factor.
RtSR [no units] is the ratio of underground area to aboveground area (Root-to-Shoot Ratio)
BD[t-d.m./m3] is Bulk Density * d.m. is dry matter
CF[t-C./t-d.m] is the carbon content
(44/12) is the conversion factor between tons of carbon and tons of carbon dioxide
BEF, RtSR, BD, and CF are taken from inventory data such as IPCC. In this calculation, data from Japan's Greenhouse Gas Inventory Report (2022) is used.

c = BEF x (1+RtSR) x BD x CF x 44/12 x$10^{-6}$
= 1.664736867 x (1+0.2715789378) x 0.3978947401 x 0.509999999905 x 44/12
= 1.575065866x$10^{-6}$

*Methodologies for calculating CO2 absorption are common among CDM, Forestry Agency, Verra, and others. When trunk volume (m3) is not measured directly, the allometry equation is commonly used.

### *2.7 Data Collection*

To verify the practicality of the above equation, the authors, with the cooperation of Gunze Green, a greening company, measured the relationship between height and diameter in a real environment. The measurement results are shown below.

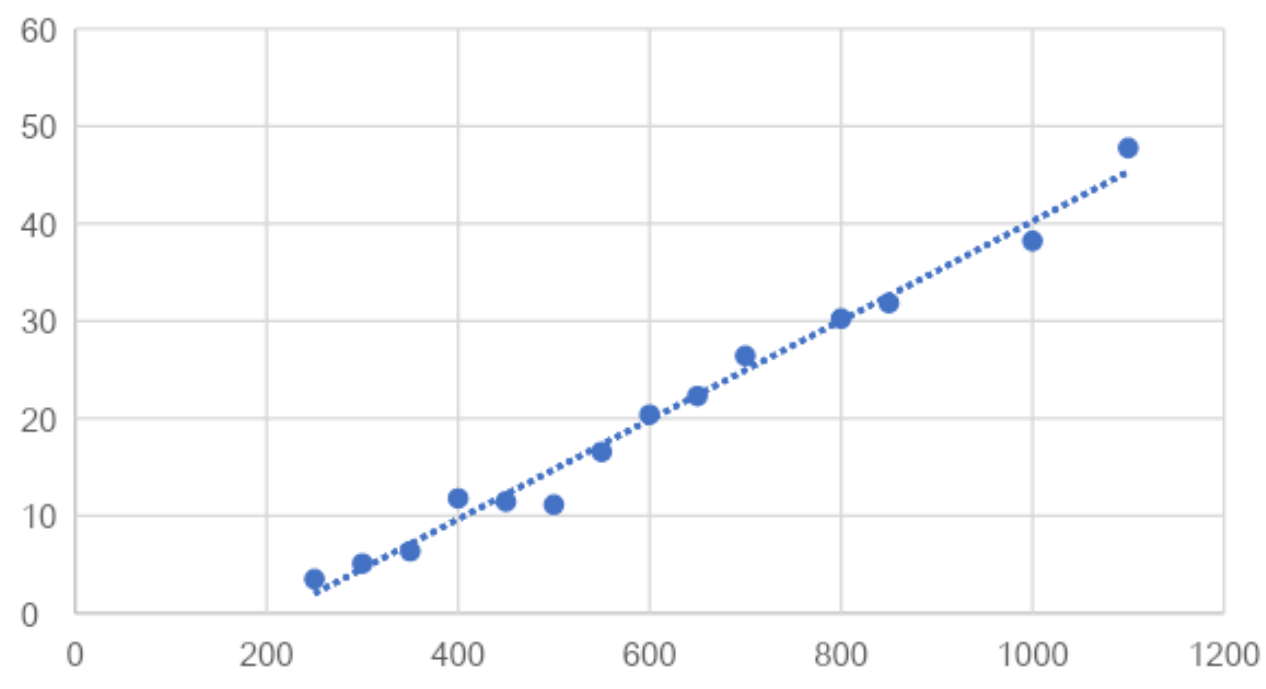

Figure 3 Evergreen, 300 cm or more but less than 1430 cm

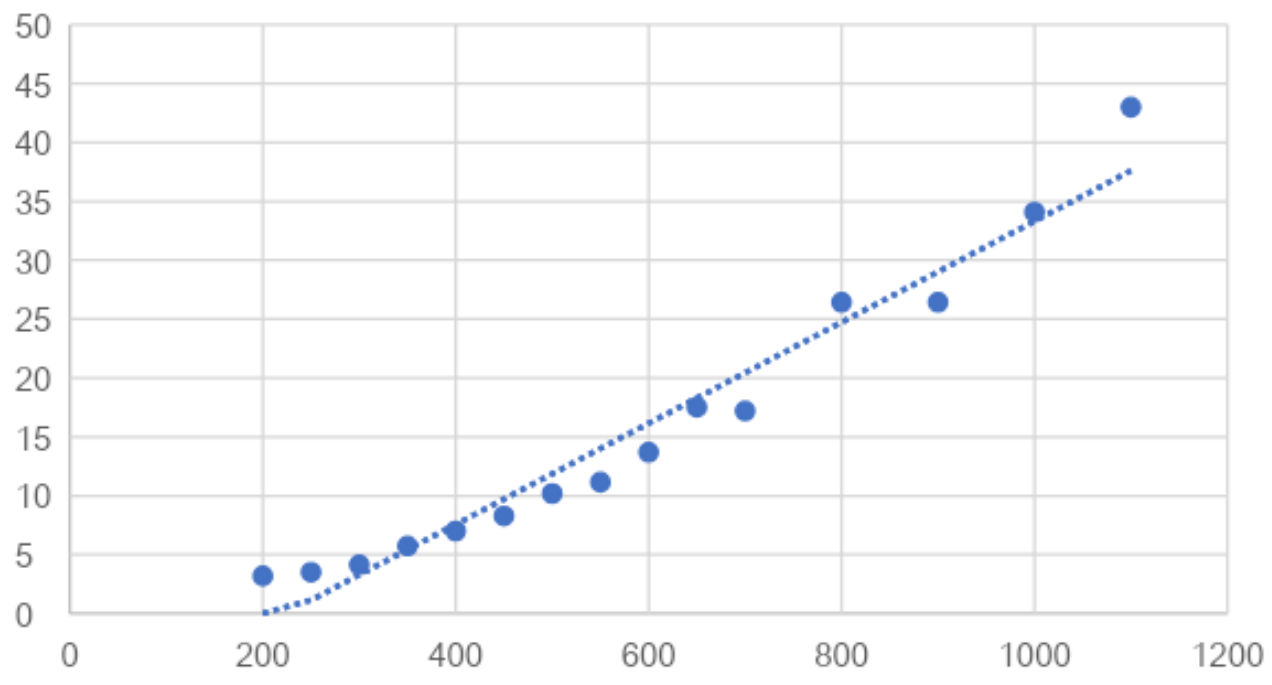

Figure 4 Deciduous tree, 300 cm or more but less than 1430 cm

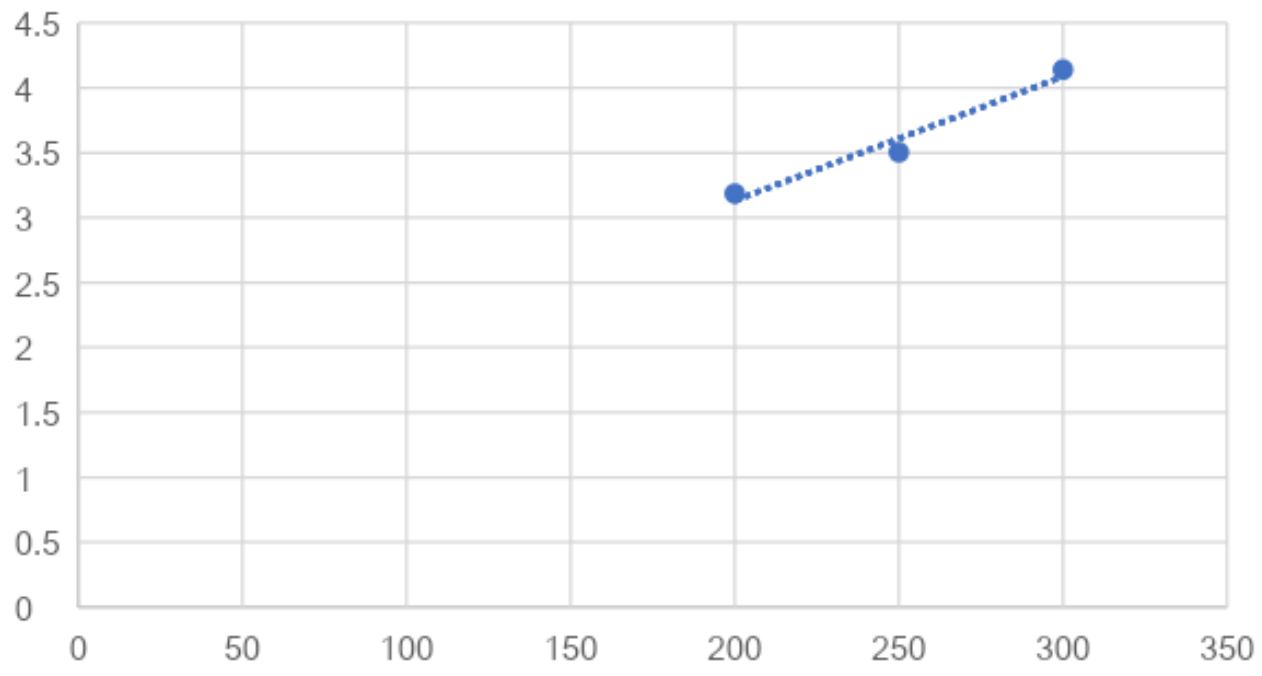

Figure 5 Deciduous tree, 0 cm to less than 300 cm

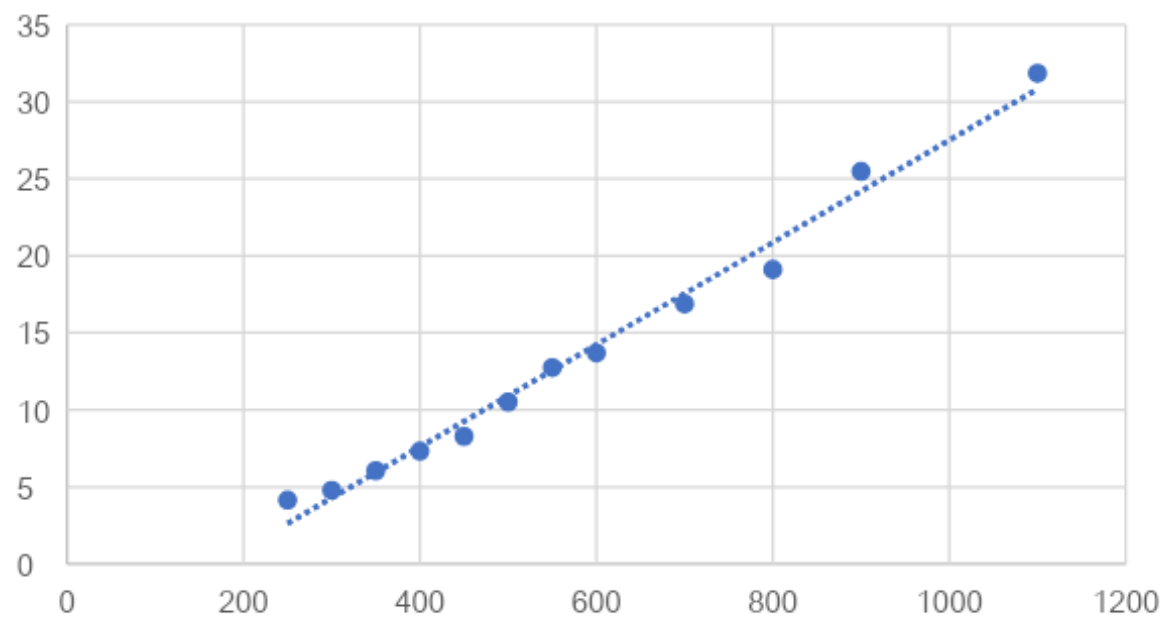

Figure 6 Conifer, 300 cm or more but less than 1430 cm

The regression equation for the above tree height-diameter relationship based on above is shown below. The validity of this equation was verified during the field survey conducted in March 2024.

Table 4 Evergreen tree equations

| Height (cm)(= H) | Diameter (cm) (= d) |
|---|---|
| 0 - 250 | d = 0.014 H |
| 250 - 300 | d = 0.0318 H - 4.4586 |
| 300 - | d = 0.051 H - 10.717 |

Table 5 Deciduous tree equations

| Height (cm)(= H) | Diameter (cm) (= d) |
|---|---|
| 0 - 300 | d = 0.0096 H + 1.2208 |
| 300 - | d = 0.0429 H - 9.5903 |

Table 6 Conifer tree equations

| Height (cm)(= H) | Diameter (cm) (= d) |
|---|---|
| 0 - 300 | d = 0.0127 H + 0.9554 |
| 300 - | d = 0.0332 H - 5.6785 |

## 3. Results

Using the calculation formulas presented above, we tested the practicality of the proposed method by performing calculations based on data actually submitted by the greening companies. The calculation results are shown below. First, the condition after 100 years by type was calculated as follows.

Table 7 Calculated parameters

| | survival rate | Height (cm) | Diameter (cm) | constant | 100-year planting absorption yield (t) |
|---|---|---|---|---|---|
| Evergreen (tall tree) | 0.062732089 | 2301.206775 | 106.6445455 | 1.57507E-06 | 2.031006398 |
| Deciduous tree (tall tree) | 0.062732089 | 2448.066545 | 95.4317548 | 1.57507E-06 | 1.730164473 |
| Conifers (tall trees) | 0.062732089 | 3242.616118 | 101.9763551 | 1.57507E-06 | 2.616814029 |
| Evergreen (medium tree) | 0.074024039 | 850 | 32.633 | 1.57507E-06 | 0.082888505 |
| Deciduous tree (medium tree) | 0.074024039 | 850 | 26.8747 | 1.57507E-06 | 0.056216986 |
| Conifer (medium tree) | 0.074024039 | 850 | 22.5415 | 1.57507E-06 | 0.039549949 |
| Evergreen (shrub) | 0.074024039 | 400 | 9.683 | 1.57507E-06 | 0.003434324 |
| Deciduous trees (shrubs) | 0.074024039 | 400 | 7.5697 | 1.57507E-06 | 0.002098837 |
| Conifers (shrubs) | 0.074024039 | 400 | 7.6015 | 1.57507E-06 | 0.002116508 |

The breakdown of CO2 absorption per tree per 100 years per period was calculated as follows.

Table 8 Breakdown for Evergreen (tall tree)

| Period since planting (t) | CO2 absorption (t) Outside of 100-year planting (trees felled or cut down in the process) | CO2 absorption (t) 100-year tree planting |
|---|---|---|
| 0 - 4.16151 | 0.000101364 | |
| 4.16151 - 5.04914 | 0.000136381 | |
| 5.04914 - 99 | 6.4738 | 2.031006398 |

Table 9 Breakdown for Deciduous tree (tall tree)

| Period since planting (t) | CO2 absorption (t) Outside of 100-year planting (trees felled or cut down in the process) | CO2 absorption (t) 100-year tree planting |
|---|---|---|
| 0 - 3.29970 | 0.00016656 | |
| 3.29970 - 99 | 7.81853 | 1.730164473 |

Table 10 Breakdown for Conifers (tall trees)

| Period since planting (t) | CO2 absorption (t) Outside of 100-year planting (trees felled or | CO2 absorption (t) 100-year tree planting |
|---|---|---|
| | cut down in the process) | |
| 1 - 2.68909 | 0.000153015 | |
| 2.68909 - 99 | 6.58251 | 2.616814029 |

Table 11 Breakdown for Evergreen (medium tree)

| Period since planting (t) | CO2 absorption (t) Outside of 100-year planting (trees felled or cut down in the process) | CO2 absorption (t) 100-year tree planting |
|---|---|---|
| 0 - 4.16151 | 0.0000959056 | |
| 4.16151 - 5.04914 | 0.00012933 | |
| 5.04914 - 16.412 | 0.0778827 | |
| 16.412 - 99 | 0.637001 | 0.082888505 |

Table 12 Breakdown for Deciduous tree (medium tree)

| Period since planting (t) | CO2 absorption (t) Outside of 100-year planting (trees felled or cut down in the process) | CO2 absorption (t) 100-year tree planting |
|---|---|---|
| 0 - 3.29970 | 0.000157376 | |
| 3.29970 - 16.412 | 0.192475 | |
| 16.412 - 99 | 0.43203 | 0.056216986 |

Table 13 Breakdown for Conifer (medium tree)

| Period since planting (t) | CO2 absorption (t) Outside of 100-year planting (trees felled or cut down in the process) | CO2 absorption (t) 100-year tree planting |
|---|---|---|
| 1 - 2.68909 | 0.000144506 | |
| 2.68909 - 16.412 | 0.132616 | |
| 16.412 - 99 | 0.303943 | 0.039549949 |

Table 14 Breakdown for Evergreen (shrub)

| Period since planting (t) | CO2 absorption (t) Outside of 100-year planting (trees felled or cut down in the process) | CO2 absorption (t) 100-year tree planting |
|---|---|---|
| 0 - 2.32558 | 0.0000539282 | |
| 2.32558 - 2.7906976744 | 0.0000714162 | |
| 2.7906976744 - 3.72093 | 0.0051921 | |
| 3.72093 - 99 | 0.0380884 | 0.003434324 |

Table 15 Breakdown for Deciduous trees (shrubs)

| Period since planting (t) | CO2 absorption (t) Outside of 100-year planting (trees felled or cut down in the process) | CO2 absorption (t) 100-year tree planting |
|---|---|---|
| 0 - 2.7906976744 | 0.000130053 | |
| 2.7906976744 - 3.72093 | 0.000303882 | |
| 3.72093 - 99 | 0.0232772 | 0.002098837 |

Table 16 Breakdown for Conifers (shrubs)

| Period since planting (t) | CO2 absorption (t) Outside of 100-year planting (trees felled or cut down in the process) | CO2 absorption (t) 100-year tree planting |
|---|---|---|
| 1 - 2.7906976744 | 0.000157871 | |
| 2.7906976744 - 3.72093 | 0.000352394 | |
| 3.72093 - 99 | 0.0234732 | 0.002116508 |

The following calculations were made for the 100-year CO2 sequestration per tree.

Table 17 100-year CO2 Sequestration per Tree

| | If only those that have survived for 100 years are accepted as credits (t) | | If any credit other than the one that lasted for 100 years is recognized as a credit (t) | |
|---|---|---|---|---|
| | total amount | Of which, contribution of greening businesses (3 years) | total amount | Of which, contribution of greening businesses (3 years) |
| Evergreen (tall tree) | 2.031006398 | 0.060930192 | 8.505044143 | 0.255151324 |
| Deciduous tree (tall tree) | 1.730164473 | 0.051904934 | 9.548861033 | 0.286465831 |
| Conifers (tall trees) | 2.616814029 | 0.078504421 | 9.199477044 | 0.275984311 |
| Evergreen (medium tree) | 0.082888505 | 0.002486655 | 0.797997441 | 0.023939923 |
| Deciduous tree (medium tree) | 0.056216986 | 0.00168651 | 0.680879362 | 0.020426381 |
| Conifer (medium tree) | 0.039549949 | 0.001186498 | 0.476253455 | 0.014287604 |
| Evergreen (shrub) | 0.003434324 | 0.00010303 | 0.046840168 | 0.001405205 |
| Deciduous trees (shrubs) | 0.002098837 | 6.29651E-05 | 0.025809972 | 0.000774299 |
| Conifers (shrubs) | 0.002116508 | 6.34952E-05 | 0.026099973 | 0.000782999 |

These series of calculation results suggest that this formula is capable of calculating carbon fixation based on field data, suggesting that the proposed method may be useful.

## 4. Discussion

In order to explore the possibility of carbon credits for greening projects, which play an important role in climate change mitigation, this paper examines a formula for estimating the amount of carbon fixation for greening activities in urban areas through tree planting.

The usefulness of the formula studied was examined by conducting calculations based on actual data through measurements made by on-site surveys of greening companies. A series of calculation results suggest that this formula may be useful.

On the other hand, the validity and accuracy of this formula have not yet been determined, and this is an issue to be addressed in the future.

Recognizing carbon credits for green businesses for the carbon sequestration of their projects is an important incentive not only as part of environmental improvement and climate change action, but also to improve the health and well-being of local communities and to generate economic benefits. This study is a pioneering exploration of the methodology. It is expected that the methodology will be refined through further research in the future.

## Conflict of Interest

This research was funded by Gunze Inc., the parent company of Gunze Green.

## Appendix

The detailed course of calculation is attached as follows.

CO2 absorption over 100 years =. $\int_0^{99} (1-p)^t pH(t)(\frac{d(t)}{2})^2 \pi c dt$ + $(1-p)^{100} H(100)(\frac{d(100)}{2})^2 \pi c$

p = 0.027309

c = 1.575065866× $10^{-6}$

Table 18 Calculation details for evergreen tree

H(t) = $2500(1-0.975^t)$

| Height (cm)(= H) | Period since planting (t) | Diameter (cm) (= d) | Outside of 100-year planting (trees felled or cut down in the process) | 100-year tree planting |
|---|---|---|---|---|
| 0 - 250 | 0 - 4.16151 | d = 0.014 H | ∮((1-0.027309)^x*0.027309*(2500*(1-0.975^x))/4*(0.014*(2500*(1-0.975^x)))^2*3.141592*1.575065866/1000000) = 0.000101364(t) | - |
| 250 - 300 | 4.16151 - 5.04914 | d = 0.0318 H - 4.4586 | ∮((1-0.027309)^x*0.027309*(2500*(1-0.975^x))/4*(0.0318*(2500*(1-0.975^x))-4.4586)^2*3.141592*1.575065866/1000000) = 0.000136381(t) | - |
| 300 - | 5.04914 - 99 | d = 0.051 H - 10.717 | ∮((1-0.027309)^x*0.027309*(2500*(1-0.975^x))/4*(0.051*(2500*(1-0.975^x))-10.717) | =(1-0.027309)^100*2500*(1-0.975^100)*(0.051*2500*(1-0.975^100)-10.717)^2*0.00 |
| | | | ^2*3.141592*1.575065866/1000000) = 6.4738(t) | 0001575065866*3.141592/4 =2.031006398 (t) |

Table 19 Calculation details for deciduous tree

$$H(t) = 2500(1 - 0.962^{t})$$

| Height (cm)(= H) | Period since planting (t) | Diameter (cm) (= d) | Outside of 100-year planting (trees felled or cut down in the process) | 100-year tree planting |
|---|---|---|---|---|
| 0 - 300 | 0 - 3.29970 | d = 0.0096 H + 1.2208 | ∮((1-0.027309)^x*0.027309*(2500*(1-0.962^x))/4*(0.0096*(2500*(1-0.962^x))+1.2208)^2*3.141592*1.575065866/1000000) = 0.00016656(t) | - |
| 300 - | 3.29970 - 99 | d = 0.0429 H - 9.5903 | ∮((1-0.027309)^x*0.027309*(2500*(1-0.962^x))/4*(0.0429*(2500*(1-0.962^x))-9.5903)^2*3.141592*1.575065866/1000000) = 7.81853 (t) | =(1-0.027309)^100*(2500*(1-0.962^100))*(0.0429*2500*(1-0.962^100))-9.5903)^2*0.000001575065866*3.141592/4 =1.730164473 (t) |

Table 20 Calculation details for conifer

$$H(t) = 35 + 5471(1 - e^{-0.00592(t-1)})^{0.65669}$$

| Height (cm)(= H) | Period since planting (t) | Diameter (cm) (= d) | Outside of 100-year planting (trees felled or cut down in the process) | 100-year tree planting |
|---|---|---|---|---|
| 0 - 300 | 1 - 2.68909 | d = 0.0127 H + 0.9554 | ∮((1-0.027309)^x*0.027309*(35+5471*(1-exp(-0.00592(x-1)))^0.65669)/4*(0.0127*(35+5471*(1-exp(-0.00592(x-1)))^0.65669)+0.9554)^2*3.141592*1.575065866/1000000) | - |
| | | | = 0.000153015(t) | |
| 300 - | 2.68909 - 99 | d = 0.0332 H - 5.6785 | ∮((1-0.027309)^x*0.027309*(35+5471*(1-exp(-0.00592(x-1)))^0.65669)/4*(0.0332*(35+5471*(1-exp(-0.00592(x-1)))^0.65669)-5.6785)^2*3.141592*1.575065866/1000000) = 6.58251 (t) | =(1-0.027309)^100*(35+5471*(1-EXP(-0.00592*(100-1)))^0.65669)*(0.0332*(35+5471*(1-EXP(-0.00592*(100-1))))^0.65669-5.6785)^2*0.000001575065866*3.141592/4 =2.616814029 (t) |

Below, the calculations were as follows:

CO2 absorption over 100 years =. $\int_0^{99} (1-p)^t pH(t)(\frac{d(t)}{2})^2\pi c dt + (1-p)^{100}H(100)(\frac{d(100)}{2})^2\pi c$

p = 0.0256977

c = 1.575065866× $10^{-6}$

Table 21 Calculation details for evergreen tree

H(t) = $2500(1-0.975^t)$

| Height (cm)(= H) | Period since planting (t) | Diameter (cm) (= d) | Outside of 100-year planting (trees felled or cut down in the process) | 100-year tree planting |
|---|---|---|---|---|
| 0 - 250 | 0 - 4.16151 | d = 0.014 H | ∮((1-0.0256977)^x*0.0256977*(2500*(1-0.975^x))/4*(0.014*(2500*(1-0.975^x)))^2*3.141592*1.575065866/1000000) = 0.0000959056(t) | - |
| 250 - 300 | 4.16151 - 5.04914 | d = 0.0318 H - 4.4586 | ∮((1-0.0256977)^x*0.0256977*(2500*(1-0.975^x))/4*(0.0318*(2500*(1-0.975^x))-4.4586)^2*3.141592*1.575065866/1000000) = 0.00012933(t) | - |
| 300 - | 5.04914 - 16.412 | d = 0.051 H - 10.717 | ∮((1-0.0256977)^x*0.0256977*(2500*(1-0.975^x))/4*(0.051*(2500*(1-0.975^x))-10.717) | |
| | | | ^2*3.141592*1.575065866/1000000) = 0.0778827(t) | |
| | 16.412 - 99 | d = 0.051 H - 10.717 | ∮((1-0.0256977)^x*0.0256977*850/4*(0.051*850-10.717)^2*3.141592*1.575065866/1000000) = 0.637001(t) | =(1-0.0256977)^100*850*(0.051*850-10.717)^2*0.000001575065866*3.141592/4 =0.082888505 (t) |

Table 22 Calculation details for deciduous tree

$H(t) = 2500(1 - 0.962^t)$

| Height (cm)(= H) | Period since planting (t) | Diameter (cm) (= d) | Outside of 100-year planting (trees felled or cut down in the process) | 100-year tree planting |
|---|---|---|---|---|
| 0 - 300 | 0 - 3.29970 | d = 0.0096 H + 1.2208 | ∮((1-0.0256977)^x*0.0256977*(2500*(1-0.962^x))/4*(0.0096*(2500*(1-0.962^x))+1.2208)^2*3.141592*1.575065866/1000000) = 0.000157376(t) | - |
| 300 - | 3.29970 - 16.412 | d = 0.0429 H - 9.5903 | ∮((1-0.0256977)^x*0.0256977*(2500*(1-0.962^x))/4*(0.0429*(2500*(1-0.962^x))-9.5903)^2*3.141592*1.575065866/1000000) = 0.192475(t) | |
| | 16.412 - 99 | | ∮((1-0.0256977)^x*0.0256977*850/4*(0.0429*850-9.5903)^2*3.141592*1.575065866/1000000) = 0.43203 (t) | =(1-0.0256977)^100*850*(0.0429*850-9.5903)^2*0.000001575065866*3.141592/4 =0.056216986 (t) |

Table 23 Calculation details for conifer

$H(t) = 35 + 5471(1 - e^{-0.00592(t-1)})^{0.65669}$

| Height (cm)(= H) | Period since planting (t) | Diameter (cm) (= d) | Outside of 100-year planting (trees | 100-year tree planting |
|---|---|---|---|---|
| | | | felled or cut down in the process) | |
| 0 - 300 | 1 - 2.68909 | d = 0.0127 H + 0.9554 | ∮((1-0.0256977)^x*0.0256977*(35+5471*(1-exp(-0.00592(x-1)))^0.65669)/4*(0.0127*(35+5471*(1-exp(-0.00592(x-1)))^0.65669)+0.9554)^2*3.141592*1.575065866/1000000) = 0.000144506(t) | - |
| 300 - | 2.68909 - 16.412 | d = 0.0332 H - 5.6785 | ∮((1-0.0256977)^x*0.0256977*(35+5471*(1-exp(-0.00592(x-1)))^0.65669)/4*(0.0332*(35+5471*(1-exp(-0.00592(x-1)))^0.65669)-5.6785)^2*3.141592*1.575065866/1000000) = 0.132616(t) | |
| | 16.412 - 99 | | ∮((1-0.0256977)^x*0.0256977*850/4*(0.0332*850-5.6785)^2*3.141592*1.575065866/1000000) = 0.303943(t) | =(1-0.0256977)^100*850*(0.0332*850-5.6785)^2*0.000001575065866*3.141592/4 =0.039549949 (t) |

Table 24 Calculation details for evergreen tree

H(t) = 107.5t

| Height (cm)(= H) | Period since planting (t) | Diameter (cm) (= d) | Outside of 100-year planting (trees felled or cut down in the process) | 100-year tree planting |
|---|---|---|---|---|
| 0 - 250 | 0 - 2.32558 | d = 0.014 H | ∮((1-0.0256977)^x*0.0256977*107.5*x/4*(0.014*107.5*x)^2*3.141592*1.5750658 | - |
| | | | 66/1000000) = 0.0000539282(t) | |
| 250 - 300 | 2.32558 - 2.7906976744 | d = 0.0318 H - 4.4586 | ∮((1-0.0256977)^x*0.0256977*107.5*x/4*(0.0318*107.5*x-4.4586)^2*3.141592*1.575065866/1000000) = 0.0000714162(t) | - |
| 300 - 400 | 2.7906976744 - 3.72093 | d = 0.051 H - 10.717 | ∮((1-0.0256977)^x*0.0256977*107.5*x/4*(0.051*107.5*x-10.717)^2*3.141592*1.575065866/1000000) = 0.0051921(t) | |
| 400 | 3.72093 - 99 | d = 0.051 H - 10.717 | ∮((1-0.0256977)^x*0.0256977*400/4*(0.051*400-10.717)^2*3.141592*1.575065866/1000000) = 0.0380884(t) | =(1-0.0256977)^100*400*(0.051*400-10.717)^2*0.000001575065866*3.141592/4 =0.003434324 (t) |

Table 25 Calculation details for deciduous tree

$H(t) = \boxed{107.5t}$

| Height (cm)(= H) | Period since planting (t) | Diameter (cm) (= d) | Outside of 100-year planting (trees felled or cut down in the process) | 100-year tree planting |
|---|---|---|---|---|
| 0 - 300 | 0 - 2.7906976744 | d = 0.0096 H + 1.2208 | ∮((1-0.0256977)^x*0.0256977*107.5*x/4*(0.0096*107.5*x+1.2208)^2*3.141592*1.575065866/1000000) = 0.000130053(t) | - |
| 300 - 400 | 2.7906976744 - 3.72093 | d = 0.0429 H - 9.5903 | ∮((1-0.0256977)^x*0.0256977*107.5*x/4*(0.0429*107.5*x-9.5903)^2*3.141592*1.575065866/1000000) = 0.000303882(t) | |
| 400 | 3.72093 - 99 | d = 0.0429 H - 9.5903 | ∮((1-0.0256977)^x*0.0256977*400/4*(0.0429*400-9.5903)^2*3.141592*1.575065866/1000000) = 0.0232772(t) | =(1-0.0256977)^100*400*(0.0429*400-9.5903)^2*0.000001575065866*3.141592/4 =0.002098837 (t) |

Table 26 Calculation details for conifer

H(t) =107.5t

| Height (cm)(= H) | Period since planting (t) | Diameter (cm) (= d) | Outside of 100-year planting (trees felled or cut down in the process) | 100-year tree planting |
|---|---|---|---|---|
| 0 - 300 | 1 - 2.7906976744 | d = 0.0127 H + 0.9554 | ∮((1-0.0256977)^x*0.0256977*107.5*x/4*(0.0127*107.5*x+0.9554)^2*3.141592*1.575065866/1000000) = 0.000157871(t) | - |
| 300 - 400 | 2.7906976744 - 3.72093 | d = 0.0332 H - 5.6785 | ∮((1-0.0256977)^x*0.0256977*107.5*x/4*(0.0332*107.5*x-5.6785)^2*3.141592*1.575065866/1000000) = 0.000352394(t) | |
| 400 | 3.72093 - 99 | d = 0.0332 H - 5.6785 | ∮((1-0.0256977)^x*0.0256977*400/4*(0.0332*400-5.6785)^2*3.141592*1.575065866/1000000) = 0.0234732(t) | =(1-0.0256977)^100*400*(0.0332*400-5.6785)^2*0.000001575065866*3.141592/4 =0.002116508 (t) |